# Hiding Single Photons With Spread Spectrum Technology

Chinmay Belthangady[1],[*] Chih-Sung Chuu[1], Ite A. Yu[2], G.Y. Yin[1], J. M. Kahn[1], and S.E. Harris[1]
[1] *Edward L. Ginzton Laboratory, Stanford University, Stanford, California 94305, USA* and
[2] *Department of Physics, National Tsing Hua University, Hsinchu 300, Taiwan.*
(Dated: April 1, 2010)

We describe a proof-of-principal experiment demonstrating the use of spread spectrum technology at the single photon level. We show how single photons with a prescribed temporal shape, in the presence of interfering noise, may be hidden and recovered.

Over the last several years scientists have learned how to extend techniques that are normally associated with classical coherent pulses of light to the single photon level. This has been motivated, in part, by the possibility of using such techniques for quantum communication and key distribution [1, 2]. As an example, it is now possible to generate single photons that have a prescribed temporal waveform [3–5], and to use high-speed electro-optic modulators to arbitrarily modulate the amplitude of single photon [6] and biphoton [7] wavepackets. Recently, by establishing a time origin with a laser pulse pumping an atom trapped in a cavity, Specht et al. [8] have demonstrated phase shaping and quantum interference of single photons. At the same time it has become clear how a single photon, based on its temporal waveform or time of arrival, may carry quantum information [9], and even, an entire image [10].

It is the intent of this Letter to further extend this classical-quantum interplay by demonstrating the use of spread spectrum technology [11] at the single photon level. We show how single photons with a prescribed temporal waveform may be transmitted through a noisy environment that is created by either narrow band thermal photons, or as in this work, an interfering laser beam that has an average power that is about a thousand times larger than the average power of the beam of single photons. We do this by using two synchronously driven electro-optic phase modulators. As shown in Fig.1(a), the modulator at the transmitter, $M_1$, broadens the spectrum of the incident photon beam from about 1 MHz to about 10 GHz and thereby reduces the spectral power density by a factor of $10^4$. The receiver modulator, $M_2$, is run in anti-phase to the transmitter modulator so as to demodulate the photon beam and reduce its bandwidth to the original 1 MHz, thereby allowing it to be transmitted through a narrow bandpass filter, $F_2$.

Now suppose that one wishes to transmit the single photon beam through an environment of thermal photons that have a linewidth that is comparable to that of the original photon beam. If this noise, or instead, a narrowband interfering laser beam is injected after the first modulator [Fig.1(a)], it is spectrally broadened by the second modulator and only a fraction of its power is transmitted through the narrow band filter at the re-ceiver. As a result, the signal-to-noise is increased approximately in the ratio of the modulation bandwidth to the bandwidth of the final filter, and in the language of spread spectrum technology the system experiences a "processing gain". In this work we demonstrate a processing gain of a factor of 50.

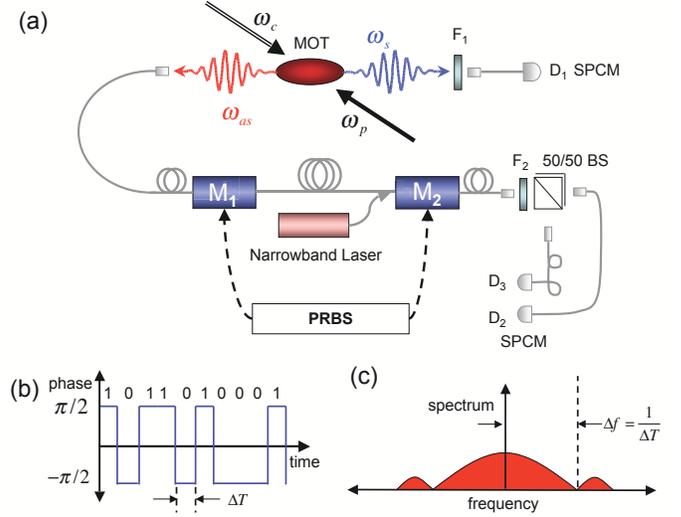

FIG. 1: (color online) Schematic of paired photon generation and spread spectrum experiment. (a) The spectrum of anti-Stokes photons produced in a magneto-optic trap is spread and narrowed by phase modulators $M_1$ and $M_2$ driven by a pseudorandom bit sequence (PRBS) generator. A weak narrowband laser is used to hide the photons. $F_1$ and $F_2$ are narrowband filters. A 50/50 beamsplitter (BS) is used to verify the presence of single photons. (b) Pseudorandom phase modulation in the time domain. The labels 1, 0 refer to the outputs of the PRBS generator. (c) Schematic of the optical spectrum after phase modulation by the PRBS generator.

In this experiment we use narrowband time-energy entangled pairs of photons (biphotons) produced in a high optical density two dimensional magneto-optic trap (MOT) comprised of laser cooled $^{87}$Rb atoms. On application of counter-propagating, continuous-wave pump, $\omega_p$ and coupling, $\omega_c$ lasers to the atom cloud, pairs of counter-propagating Stokes, $\omega_s$ and anti-Stokes, $\omega_{as}$ photons are spontaneously generated in the medium and col-lected into single-mode fibers. Because of electromagnetically induced transparency (EIT), the coupling laser renders the atom cloud transparent at the anti-Stokes frequency. As a result of the slow light effect associated with the narrow EIT window, the anti-Stokes photon propagates with a group velocity that is about $10^4$ times slower than the Stokes photon, and the temporal length of the biphoton wavepacket (250 ns) is approximately equal to the dimension of the Rb cloud (1.7 cm) divided by group velocity of the anti-Stokes photon [12]. The physics of the paired photon generation process has been discussed by Balic et al. [4], and Kolchin [13], and operation with a subnatural linewidth has have been reported by Du et al. [14].

We set the optical depth of the MOT at 30. With $2\gamma_{13}$ equal to the spontaneous decay rate out of state $|3\rangle$, we set the coupling laser Rabi frequency $\Omega_c = 4.01\gamma_{13}$, the pump laser Rabi frequency $\Omega_p = 1.66\gamma_{13}$ and the pump laser detuning, $\Delta\omega_p = 48.67\gamma_{13}$. Under these conditions the source produces narrowband biphotons with an estimated linewidth of 3.5 MHz at a rate of about 11500 pairs/s.

The detection of the Stokes photon by detector $D_1$ sets the time origin for measurement of the biphoton wavefunction at detector $D_2$. The anti-Stokes photon is sent through a 20-GHz electro-optic phase modulator (Eospace Inc.) driven by one of two differential data outputs of a pseudorandom bit sequence (PRBS) generator (Centellax TG2P1A) with a bit rate tunable between 0.05-10 Gb/s. In each time slot $\Delta T$, equal to the inverse bit rate, the PRBS outputs a random voltage equal to $V_\pi/2$ (corresponding to logic 1) or $-V_\pi/2$ (logic 0), where $V_\pi$ is the half-wave voltage of the phase modulators. We choose a bit pattern consisting of 32767 ($2^{15}-1$) random bits. At a bit rate of 10 Gb/s, the phase therefore switches randomly [Fig.1 (b)] between $-\pi/2$ and $\pi/2$ radians every 0.1 ns. This process of random phase modulation in the time domain spreads the spectrum in the frequency domain from about 3.5 MHz for the unmodulated photon to about 10 GHz for the modulated photon. The envelope of the spread spectrum is a sinc-squared function with the first null, as shown in Fig.1 (c), at a frequency equal to the bit rate.

The phase-modulated anti-Stokes photon is transmitted over a channel consisting of a long single-mode optical fiber [Fig. 1(a)]. At the receiver end, the photon is sent through a second phase modulator, identical to the first and driven by the complementary output of the PRBS generator. The two phase modulators are synchronized by adjusting the lengths of the electric cables connecting them to the PRBS generator such that the relative time delay, $\Delta t_{RF}$ of the radio frequency signal sent to the two modulators equals the travel time, $\Delta t_p$, of the anti-Stokes photon between the modulators. Fine adjustment of $\Delta t_{RF}$ is achieved by means of a radio frequency waveguide of variable length connected to one of the two rf cables. When the phase modulators are run in synchronism such that the phase imposed on the photon by the first modulator is exactly undone by the second modulator, the spectral width of the photon is restored to its original value of 3.5 MHz. When the phase modulators are not run in exact phase opposition, the uncompensated residual phase results in a spectrum that is wider than 3.5 MHz. After passing through the second phase modulator the anti-Stokes photon is sent through a 65-MHz fiber based Fabry-Perot filter (Micron Optics) with a free spectral range of 13.6 GHz. If the spectral width of the photon after the second phase modulator is less than the filter bandwidth, the photon passes through with little loss of probability amplitude. Conversely, if the spectral width of the photon is larger, then it is attenuated by the filter. The spectrum of narrowband noise entering the channel after the first modulator is spread by the second modulator and is attenuated by the filter. To hide the photon we use an attenuated beam from an external-cavity diode laser with a linewidth of about 300 kHz set at the same frequency as the anti-Stokes photon. A 50/50 beamsplitter (BS) and a third detector $D_3$ is placed after the filter to verify the single photon nature of the experiment.

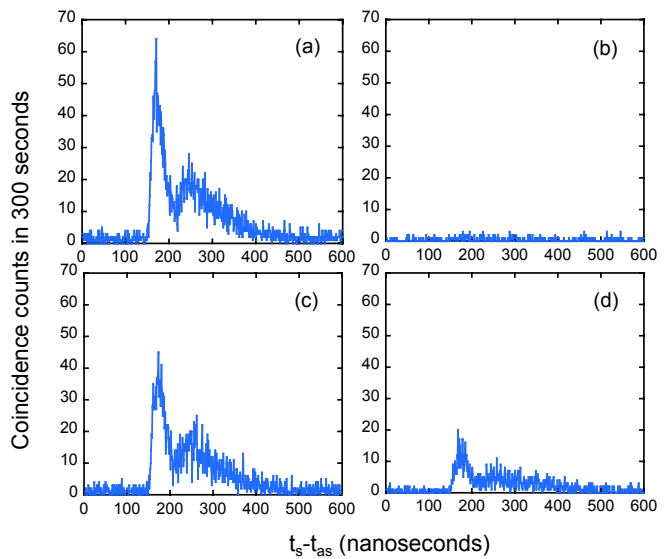

FIG. 2: (color online) Shape of the anti-Stokes photon wavepacket measured by recording coincidence counts between detectors $D_1$ and $D_2$ with the beamsplitter removed. (a) Anti-Stokes photon wavepacket measured when modulators $M_1$ and $M_2$ are off. (b) Modulator $M_1$ is on and modulator $M_2$ is off. (c) Both modulators are on and run synchronously. (d) Both modulators are on with $\Delta t_{RF} > \Delta t_p$ (see text).

With both phase modulators and the (noise simulating) laser turned off, the shape of the anti-Stokes photon wavepacket is shown in Fig. 2(a). This shape is measured by recording coincidence counts between detectors $D_1$ and $D_2$, with the beamsplitter removed, as a function of the difference in arrival times of the Stokes, $t_s$, and anti-Stokes, $t_{as}$, photons at the two detectors. Coincidence measurement is done by means of a time-to-digital converter (FAST Comtec TDC 7886s) with counts binned into 1-ns bins. The waveform consists of a sharp leading edge spike called a Sommerfield-Brillouin precursor [15], and a main waveform whose width is approximately equal to the group delay at the anti-Stokes frequency [14]. Conditioned on the detection of the Stokes photon, and neglecting the flat uncorrelated background, this shape represents the squared modulus of the wavefunction of the anti-Stokes photon. When the first phase modulator is switched on, the spectrum of this photon is spread to about 10 GHz. If the second phase modulator is off, the anti-Stokes photon is severely attenuated by the narrowband filter $F_2$ and very little photon probability amplitude leaks through to detector $D_2$. This is shown in Fig.2 (b). When the second phase modulator is switched on and driven so as to cancel the phase imposed by the first modulator, the photon spectrum is narrowed and passes through filter $F_2$ with little loss of probability amplitude. This is shown in Fig.2(c). The small residual loss in photon probability amplitude probably arises from a slight mismatch between the two modulators and their complementary driving signals. In Fig.2(d), the difference in lengths of the electric cables connecting the PRBS to the two modulators is chosen such that $\Delta t_{RF}$ is greater than $\Delta t_p$ by about 40ps. As a result, the phase imposed on the photon by the first modulator is not exactly negated by the second modulator. Now, after transmission through filter $F_2$ we observe a reduced, but not zero, coincidence count rate. A similar result is obtained when $\Delta t_{RF} < \Delta t_p$.

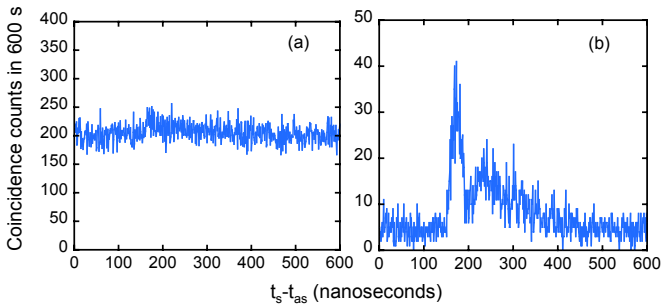

FIG. 3: (color online)Retrieval of the anti-Stokes photon from injected noise. (a) Anti-Stokes photon wavepacket as hidden by an additional laser beam that is injected into the channel. Here, both modulators are off. (b) Switching on both modulators and running them synchronously recovers the anti-Stokes photon (see text).

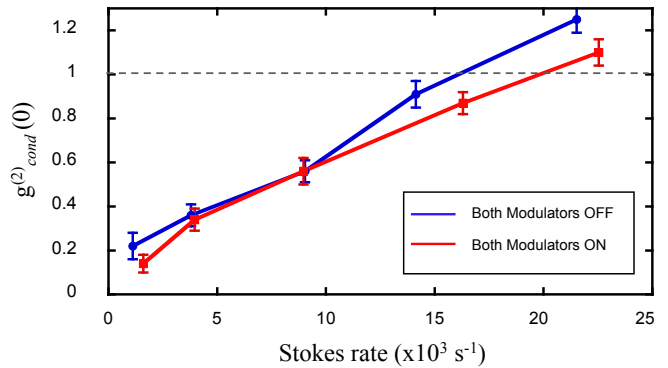

FIG. 4: (color online) Conditional threefold correlation function as a function of Stokes rate with both modulators off (blue) and with both modulators on (red). The dashed line is the classical limit. The errors are calculated by assuming a Poisson distribution for the coincidence rates.

We next deliberately inject a weak diode laser beam to hide the the anti-Stokes photon. With additional laser photons at a rate of 40000 $s^{-1}$ and anti-Stokes signal photons at a rate of 30 $s^{-1}$ [Fig. 3 (a)], anti-Stokes photons can no longer be seen. If the two modulators are now turned on and run in synchronism, the spectrum of the anti-Stokes photon is spread by the first modulator and compressed by the second modulator so that the photon passes through the filter and arrives at the detector with very little attenuation. The spectrum of the laser photons, on the other hand, is spread by the second phase modulator and attenuated by the filter and very little of the laser power reaches the detector. When this is the case, the shape of the anti-Stokes photon wavepacket can be recovered from the noise [Fig.3(b)]. The decrease in the level of the flat uncorrelated background by a factor of about 50 corresponds to the spread spectrum processing gain.

We next test to insure that the experiment results are dominated by single photon events. To the extent that the down conversion process produces only biphotons, and when the square of the rate of biphoton generation multiplied by the gate width is much less than unity, then no threefold coincidences will be observed between the detectors $D_1$, $D_2$ and $D_3$. The quality of a single photon source can be quantified by means of the conditional Glauber correlation function [6] defined as

$$g^{(2)}_{cond}(0) = \frac{N_{123}N_1}{N_{12}N_{13}} \quad (1)$$

where $N_{123}$ is the number of threefold coincidences between detectors $D_1$, $D_2$ and $D_3$; $N_{12}, N_{13}$ are the number of twofold coincidences between detectors $D_1, D_2$ and $D_1, D_3$, respectively; and $N_1$ is the number of counts detected by detector $D_1$. Here all coincidence counts are measured within a time window equal to the nominal length of the anti-Stokes photon wavepacket. We turn off the the weak narrowband laser beam and measure $g^{(2)}_{cond}(0)$ as a function of the rate of detected Stokes photons with both modulators turned off (blue line in Fig. 4) and with both modulators turned on and running synchronously (red line in Fig. 4). At a Stokes rate of approximately $1200 s^{-1}$ we obtain $g^{(2)}_{cond}(0) = 0.22 \pm 0.06$ for the unmodulated photon and $g^{(2)}_{cond}(0) = 0.14 \pm 0.04$ for the phase modulated-demodulated photon. These values of $g^{(2)}_{cond}(0)$ are less than the classical limit of 1 (dashed line in Fig. 4) and reasonably less than the limiting value of 0.5; i.e., the conditional Glauber correlation function for a two-photon Fock state.

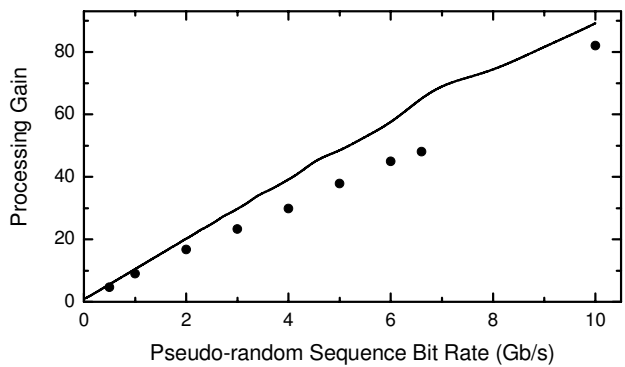

FIG. 5: Spread spectrum processing gain as a function of the PRBS bit rate. The measurement is made with a cw diode laser. Dots represent experimental data and the solid curve is calculated (see text).

As an additional check on the modulators that are used in this work, as shown in Fig. 5, we measure the transmission of the final filter versus the bit rate of the PRBS source. We send a laser beam with a linewidth of about 300 kHz through phase modulator $M_2$ and narrowband filter $F_2$. The ratio of the laser power detected after the filter with the modulator off to the power measured with the modulator on gives the value of what is termed as the spread spectrum processing gain. The dots are experimental data and the solid line is calculated by evaluating the overlap integral between the transfer function of the Fabry-Perot filter and the sinc squared spectrum produced by pseudorandom phase modulation. The fact that the experimental points lie below the calculated curve indicates that the spectrum of the laser beam is not spread to an ideal sinc-squared spectrum with the first null at a frequency equal to the bit rate. This deviation tends to be larger for higher bit rates, suggesting that it arises from the finite bandwidth of the modulators and the drive electronics.

This work has shown how narrowband single photons may be phase modulated so as to increase the width of their spectrum by several orders of magnitude while at the same time retaining the information that characterizes their waveform. By compensation with an antiphased modulator, the biphoton waveform may be reconstructed at a distant location. We have also shown how spread spectrum technology may be used to hide a single photon in the presence of laser photons of the same frequency and similar linewidth. Applications may include an additional level of classical security for quantum key distribution, and someday, multiplexing of single photon communication channels. This work may also be the first use of spread spectrum techniques combined with single photon detection.

This work was supported by the U.S. Air Force Office of Scientific Research, the U.S. Army Research Office, and the Defense Advanced Research Projects Agency.